# Percolative Charge Transport In Binary Nanocrystal Solids

*Luman Qu, Davis Unruh\*, and Gergely T. Zimanyi*

Physics Department, University of California, Davis, Davis, California 95616, United States

## ABSTRACT

We simulated electron transport across a binary nanocrystal solid (BNS) of PbSe NCs with diameters of 6.5nm and 5.1nm. We used our Hierarchical Nanoparticle Transport Simulator HINTS to model the transport in these BNSs. The mobility exhibits a minimum at a Large-NC-fraction $f_{LNC}$=0.25. The mobility minimum is deep at T=80K and partially smoothed at T=300K. We explain this minimum as follows. As the LNC fraction $f_{LNC}$ starts growing from zero, the few LNCs act as traps for the electrons traversing the BNS because their relevant energy level is lower. Therefore, increasing the $f_{LNC}$ concentration of these traps decreases the mobility. As increasing $f_{LNC}$ reaches the percolation threshold $f_{LNC}$=$f_p$, the LNCs form sample-spanning networks that enable electrons to traverse the entire BNS via these percolating LNC networks. Transport through the growing percolating LNC networks drives the rapid growth of the mobility as $f_{LNC}$ grows past $f_p$. Therefore, the electron mobility exhibits a pronounced minimum as a function of $f_{LNC}$, centered at $f_{LNC}$=$f_p$. The position of the mobility minimum shifts to larger LNC fractions as the electron density increases. We have studied the trends of this mobility minimum with temperature, electron density, charging energy, ligand length, and disorder. We account for the trends by a "renormalized trap model", in which capturing an electron renormalizes a deep LNC trap into a shallow trap or a kinetic obstacle, depending on the charging energy. We verified this physical picture by constructing and analyzing heat maps of the mobile electrons in the BNS.

## Introduction

Colloidal semiconductor nanocrystals (NCs) are exciting nanoscale building blocks for fabricating mesoscale materials that exhibit emergent collective properties. NCs are well-defined building blocks that can be synthesized with excellent control of composition, size, and shape. The energetics and the charge transport in NC solids can be tuned by changing the NC size, size distribution, shape, inter-NC spacing, spatial ordering, surface chemistry and defects, and the properties of the matrix between the NCs. This remarkable tunability makes NC solids promising promising platforms for optoelectronic applications,[1,2] including third generation solar cells,[3,4] light emitting diodes,[5] and field effect transistors (FETs).[6,7]

NC solids are especially interesting for solar cell applications because the band gap can be tuned by changing the NC size in order to improve device power conversion efficiency. Solar cell efficiency can also be improved by leveraging quantum confinement to open new energy conversion channels such as the down-converting carrier multiplication (CM), in which more than one electron-hole pairs are generated per absorbed photon.[3,8-11] CM has the potential to boost solar cell efficiency to ~44%, well beyond the Shockley-Queisser limit of ~33%.[12] Very recently, we have advocated for the formation of mini-bands in NC solar cells to implement the complementary, up-converting intermediate band solar cell paradigm.[13] In principle, intermediate band solar cells can deliver up to 47% efficiency at one sun illumination.

One of the factors limiting the utility of NC-based optoelectronics is the relatively high energetic and spatial disorder of NC solids. This disorder causes decoherence of the electronic wave functions between NCs and inhibits the emergence of new collective mesoscale behavior, resulting instead in weakly-coupled NC films with slow hopping transport. These factors, weak coupling and slow transport are the primary agents hindering the realization of high-performance NC optoelectronics.[14-16] These factors used to limit the hopping mobility in NC solids into the $10^{-4} - 10^{-1}$ cm²/Vs range. Recently, new reasons for optimism emerged as various groups managed to boost the mobility by increasing the inter-NC charge transfer rate with a variety of methods, including ligand engineering,[17-19] band alignment engineering,[20] chemical doping,[21,22] photo-doping,[23] metal-NC substitution,[24] epitaxial attachment of NCs,[16,25] and atomic layer deposition (ALD) infilling.[26] In some cases, these efforts managed to reach mobilities exceeding 10 cm²/Vs.[27]

To increase the mobility even further, a deeper understanding of the role of disorder is essential. A particularly promising testing platform is the class of binary nanocrystal solids (BNSs), which are crystalline solids composed of two different types of NCs. These metamaterials can be formed from NCs of different composition and/or size.[1,16,28-31] The Murray group demonstrated the possibility of fabricating such binary nanocrystal solids as large area monolayer and bilayer structures.[29] Later, they were able to perform an in situ ligand exchange, producing ultrafast directional carrier transfer on the timescale of 1 ps.[30] They were also able to increase the conductivity by more than three orders of magnitude by substituting an increasing fraction of PbS NCs with Ag or Au NCs.[24] The Alivisatos group focused on the percolation aspects of the electron transport, and beautifully imaged charge percolation pathways.[28] Whitham et al. devised ingenious ways to extract the localization length of the electrons in a type of percolative NC systems to further characterize transport.[16]

Theoretical efforts have kept pace with these promising experimental developments only partially. A pioneering study of charge transport in NC arrays was performed by Chandler and Nelson.[32] They modeled the electronic structure of individual NCs using the $k \cdot p$ method, then performed Monte Carlo transport studies on small samples of 2x2x3 and 3x3x4 NCs. In the case of low charging energy, metal-insulator transitions were observed at electron occupation levels $\langle n \rangle$ that corresponded to the complete filling of an s, p, or d shell. When the charging energy became comparable to the level broadening, additional minima appeared in the conductance at every integer value of $\langle n \rangle$ as a result of electron-electron repulsion. The charge transport properties of NCs embedded in a matrix were explored by others using a kinetic Monte Carlo (KMC) method.[33,34] Although these papers developed an advanced method that was capable of handling the long-range Coulomb interactions, the small system sizes limited the definitiveness of their conclusions.

Our first contribution to this field was to develop the kinetic Monte Carlo platform HINTS – the **Hi**erarchical **Na**noparticle **T**ransport **S**imulator – to compute the electron and hole mobilities as a function of the NC diameter.[35] We found that the mobility exhibited a maximum or plateau as a function of the NC diameter (depending on the type of disorder). This finding was in agreement with corresponding experiments.[36,37] Since then, we extended HINTS to describe the metal-insulator transition in NC solids and proposed a quantum percolation model to explain the unique criticality observed.[38, see also 39] Very recently, we demonstrated that NC solids are an excellent platform to study Mott-Hubbard phenomena, as they exhibit transport transitions driven by interactions, by disorder, and by their interplay.[40]

In this paper, we adapt and apply our hierarchical transport simulator HINTS[35,38] to study charge transport in binary nanocrystal solids (BNSs) consisting of PbSe NCs of two different sizes. Our simulation results are motivated by early stage efforts to measure the transport of such binary NC films fabricated with layer-by-layer dip coating of colloidal mixtures. Our main results include the following. First, our HINTS simulations of the mobility of field-effect transistors made from PbSe NCs with a mixture of 6.5 nm diameter large NCs (LNCs), and 5.1 nm diameter small NCs (SNCs) showed a deep minimum at a fraction of the LNCs, $f_{LNC}$=0.25. The minimum persists up to temperatures where $kT$ becomes comparable to the difference of the conduction band edge energies of the LNCs and SNCs. We developed a percolative theory to explain this deep mobility minimum. We propose that at low $f_{LNC}$, the LNCs form traps and thus suppress the mobility. With increasing $f_{LNC}$, these LNC traps coalesce into a percolative transport pathway at $f_{LNC} = f_p$. $f_{LNC}$ increasing beyond $f_p$ opens a new transport channel: the electrons propagating through the percolating LNCs. This new channel starts boosting transport with increasing $f_{LNC}$, thereby explaining the mobility minimum. Second, we analyzed the impact of NC site energy disorder, carrier density, charging energy, and ligand length on the transport. We developed an electron-occupation-induced trap renormalization model that accounted for the dependence of the mobility on these four parameters. Finally, we validated our percolative theory of LNC traps transforming into percolative transport pathways as a function of $f_{LNC}$ by constructing and analyzing heat maps of the carrier residence times.

## Overview of Simulation and Results

For this work, we extended and adapted our previously developed HINTS to describe transport in binary NC solids. The presentation and discussion of our results requires a brief description of the hierarchical levels of HINTS (described in more detail in the Supplemental Material).

(1) We used the event-driven molecular dynamics code PackLSD[41] to generate a random-packed, jammed NC solid for the simulation. The NC solids typically included many hundred NCs with a form factor of 10×10×1, inspired by the experimental geometry of 2D FET channels. For example, a monodisperse sample containing 400 NCs was packed into a simulation volume with the approximate spatial extent of 16×16×1.6 NC diameters. For the binary NC solids, the diameters of the small NCs (SNCs) and large NCs (LNCs) were selected from corresponding Gaussian distributions, with widths $\sigma_{SNC}$ and $\sigma_{LNC}$, and then jam-packed to form the binary NC solid BNS. We note that experimentally the amount of disorder in the NC solid can be tuned, from an approximately ordered solid to a more disordered solid, with some degree of glassiness.

(2) Next, the energy parameters of the Hamiltonian of each NC were established as follows. We used the photoelectron spectroscopy results of Jasieniak *et al*.[42] modified by the method of Miller *et al*.[43] to estimate the energies of the valence band maximum ($E_{VBM}$) and conduction band minimum ($E_{CBM}$) of PbSe NCs as a function of NC diameter. These energies are modified, or tuned, from their bulk values by "quantum confinement", the fact that the electron wavefunctions are localized on the NCs. The values are plotted in Figure 1. Photoelectron spectroscopy provides a direct measurement of $E_{VBM}$, from which $E_{CBM}$ can be estimated by adding the NC band gap. Following Jasieniak *et al*., we show limiting values of $E_{CBM}$ obtained using the experimentally-determined optical band gap ($E_{CBM,optical}$) and a calculated upper estimate of the electronic band gap, which includes Coulomb and polarization energies ($E_{CBM,max}$), yielding a range of $E_{CBM}$ values for each NC size (green band in Fig. 1). Clearly, $E_{CBM}$ exhibits a much larger change with NC size than does $E_{VBM}$. For example, 75-85% of the difference in band gap for 5.1 nm and 6.5 nm NCs is due to the change in $E_{CBM}$, with the exact value depending on the true electronic band gap of the PbSe NC. Building on these experimental findings, for our HINTS simulations we used the $E_{CBM}$ curve from the **k·p** calculations of Kang and Wise (dashed line in Fig. 1),[44] which is well-validated by its dependence on the NC size closely tracking the experimental $E_{CBM,max}$ values.

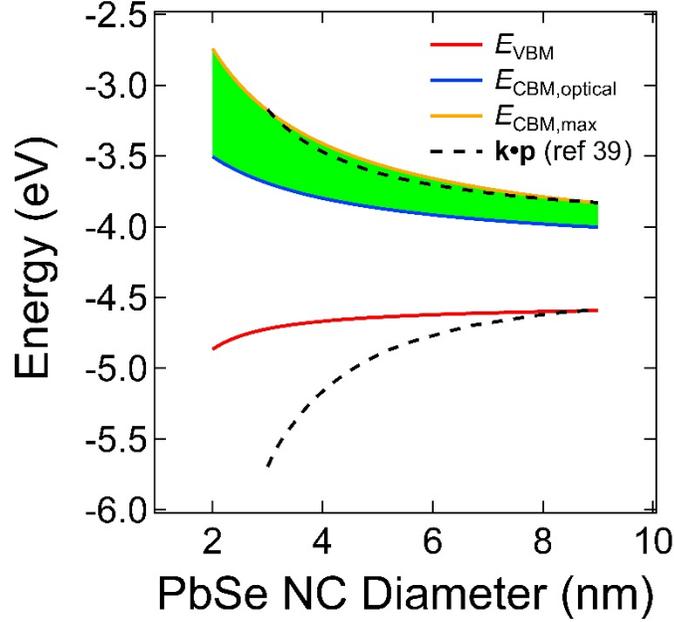

**Figure 1.** Absolute conduction and valence band edge energies ($E_{CBM}$ and $E_{VBM}$) of PbSe NCs as a function of NC diameter. $E_{VBM}$ was measured by photoelectron spectroscopy.[42,43] $E_{CBM}$ has a range of values (green band) bracketed by $E_{CBM,optical}$ (from the measured optical band gap) and $E_{CBM,max}$ (from a DFT-calculated upper estimate of the electronic band gap). Also shown are the $E_{CBM}$ and $E_{VBM}$ curves calculated from **k·p** theory (dashed lines, ref. 44), rigidly shifted along the ordinate to match the experimental data at large NC size. The latter $E_{CBM}$ curve is used to parameterize our transport model. Contribution to this Figure by Matt Law is gratefully acknowledged.

Our model also includes the electron-electron interaction at the level of an on-site self-charging energy $E_c$. This charging energy can be calculated by a variety of methods, including the semi-empirical pseudopotential configuration interaction (CI) method of Zunger and coworkers,[45,46] and the tight-binding many body perturbation theory method of Delerue.[47, 48] In this paper, we report results with the latter approach, selected for its versatility. The long-range part of the Coulomb interaction plays a noticeable part only at low temperatures, and thus can be disregarded for many applications such as solar cells.

(3) We modelled the transport of charge carriers between adjacent NCs as Miller-Abrahams phonon-assisted tunneling, also called thermally activated hopping. The inter-NC distance was set at twice the ligand length. The tunneling probabilities between NCs were determined by the standard WKB form.

(4) We simulated the transport across the BNS using the extended kinetic Monte Carlo method with activated hopping transitions between nearest-neighbor NCs.[35,49] Electrons were added to the BNS at random with a specified average electron density (e⁻/NC). Then the electrons were allowed to relax to favorable energy configurations. Transport was simulated by applying an electric field and measuring the responsive current. The simulation was run for a sufficiently long time to ensure that the carrier flow reached a steady state. The electric fields were kept sufficiently small to ensure that the transport was in the linear response regime. The carrier mobility was extracted from the slope of the linear I-V curves. The overall NC-NC hopping attempt rate prefactor was selected such that the simulated mobilities were consistent with published experimental values of the Law group.[36] We systematically explored wide ranges of temperature, disorder, electron density, and Coulomb interaction. The mobilities were determined by simulating at least 40 samples, and often more than 100 samples, at each set of parameters and then averaging the results. Since each mobility was determined from a dynamic flow of hundreds of electrons over $10^5$-$10^6$ time steps, we achieved a remarkably good self-averaging with small error bars even with this moderate number of samples.

Next, we describe the experimental motivation for our work, the main simulation results obtained using HINTS, and the physical picture emerging from the simulations.

**Fig. 2a** illustrates charge transport in a binary PbSe NC solid. Quantum confinement modifies the $1S_e$-$1S_h$ band gaps of the LNCs to be smaller than those of the SNCs, and the $1S_e$ energies (i.e. $E_{CBM}$) of the LNCs lower than those of the SNCs. Therefore, in a binary NC solid with a small $f_{LNC}$ fraction, the LNCs act as traps for the mobile electrons. (**Fig. 2a**, top). Accordingly, as $f_{LNC}$ is increased from small values, the increasing trap density causes a monotonic decrease of the carrier mobility.

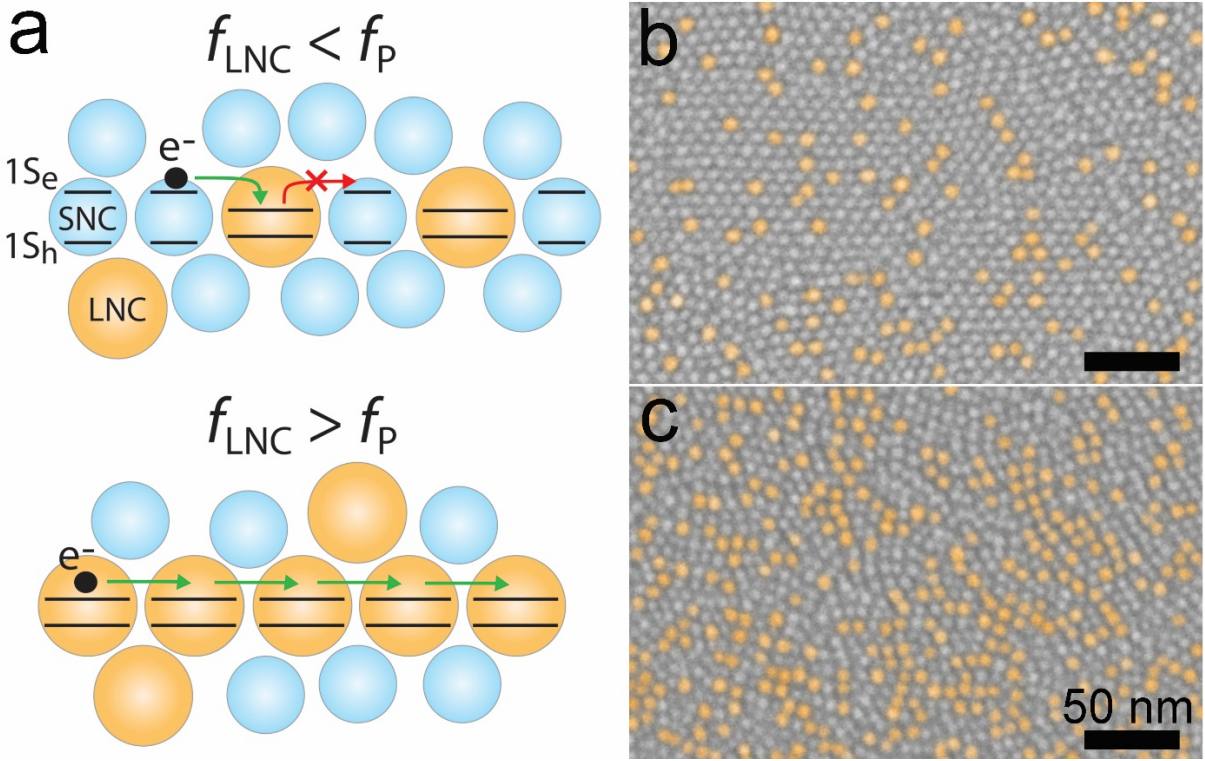

**Fig. 2a** Schematic representation of a binary NC solid containing (*top*) a small fraction of LNCs and (*bottom*) a larger fraction of LNCs. When $f_{LNC} < f_P$, the LNCs act as isolated carrier traps, impeding transport. When $f_{LNC} > f_P$, the LNCs form contiguous low-energy transport pathways that facilitate transport. **2b-2c** Colorized SEM images of a monolayer of a binary NC solid made from 6.5 nm and 5.1 nm PbSe NCs with **2b** $f_{LNC}$ = 0.12 ($< f_P$) and (**c**) $f_{LNC}$ = 0.31 ($> f_P$). The LNCs are shaded orange. LNC percolation pathways are evident in the latter image. The NCs are capped by oleate ligands (prior to exchange with EDT). Scale bars are 50 nm. SEM image is courtesy of Matt Law, UC Irvine.

Increasing $f_{LNC}$ first enables the LNCs to form clusters, and then increases the size of the LNC clusters in the BNS. Once $f_{LNC}$ exceeds the percolation threshold $f_P$, the LNC clusters interconnect to form contiguous percolation networks that span the entire BNS. These LNC percolation networks open new transport channels: low-energy, low-disorder and thus high-mobility transport pathways (**Fig. 2a**, bottom). As $f_{LNC} \rightarrow 1$, the percolation networks densify and the mobility steadily increases toward its value in a monodisperse NC solid, formed only from LNCs.

**Figs. 2b-2c** compare scanning electron microscopic images of monolayer-thick films of oleate-capped 6.5 nm and 5.1 nm PbSe NCs having $f_{LNC}$ fractions below and above $f_P$. For $f_{LNC}$ = 0.31 ($> f_P$), the LNCs

(colored orange) visibly form essentially sample-spanning clusters. These images show that the LNCs are well mixed in the SNC films and have no tendency to phase separate into pure LNC domains.

Before proceeding, a word about the experimental systems. The binary NC films were made by layer-by-layer dip coating of colloidal mixtures of 6.5 nm and 5.1 nm PbSe NCs. The NCs were synthesized by the hot injection method. Colloidal solutions of different LNC number fraction ($f_{LNC}$) were prepared in hexane, dip coated onto prepatterned field-effect transistor (FET) substrates using ligand exchange with 1,2-ethanedithiol (EDT), and then infilled and overcoated with amorphous alumina using low-temperature atomic layer deposition (ALD), yielding transistors with dominant $n$-channel (electron) transport, excellent stability, and greatly-reduced $I$-$V$ hysteresis compared to EDT-treated NC FETs before ALD infilling.[50] The $f_{LNC}$ values of the resulting films were confirmed by analyzing SEM images of the first NC monolayer in the FET channel (like those in **Figs. 2b-2c**). The experimental methods are described in further details in the Supporting Information. We continue with presenting the results of our simulations.

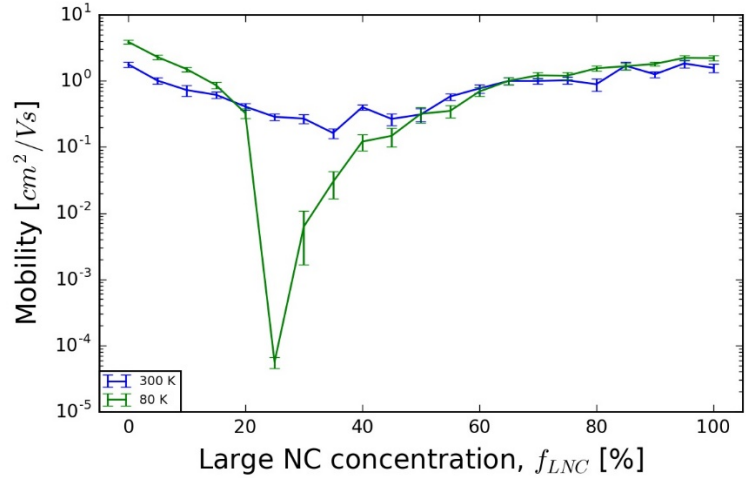

**Fig. 3.** Simulated electron mobility versus $f_{LNC}$ for 6.5 nm and 5.1 nm PbSe NCs. $\Delta E_{CBM} = 60$ meV; with a diameter disorder of $\sigma_{LNC} = 0.325$ nm and $\sigma_{SNC} = 0.046$ nm; and electron density $n = 0.25$ $e$/NC.

**1. Mobility as a function of LNC fraction $f_{LNC}$:** **Fig. 3** presents the HINTS-simulated mobility of a PbSe BNS with NC diameters of 6.5 nm and 5.1 nm at $T = 80$ K and 300 K. The corresponding difference of the SNC and LNC conduction band energy minima was taken from **Fig. 1** as $\Delta E_{CBM}$=60 meV; for other parameters see the Supporting Information. The simulated mobility curve at $T$=80 K shows a deep mobility minimum at $f_{LNC}$=0.25. Interpreting $f_{LNC}$ of this minimum as approximately equal to $f_P$, the geometric percolation fraction is consistent with the prediction of $f_p$=0.24 of bond percolation theory,[51] as well as with the percolation onset of $f_P$=0.22, observed in our recent calculations of NC films.[37] The precise relationship between the position of the mobility minimum and $f_P$ is impacted by the NC packing and the electron density, as discussed below.

The key message of this simulation is the emergence of a mobility minimum as a function of $f_{LNC}$ around $f_{LNC}$=$f_p$, which is sharp and deep at low temperatures, and partially smoothed at higher temperatures. As mentioned earlier, we developed a percolative theory to explain this deep mobility minimum. We propose that as the lower-energy LNCs are introduced into the higher-energy SNC matrix at low $f_{LNC}$, the LNCs form traps and thus suppress the electron mobility. With increasing $f_{LNC}$, these LNC traps coalesce into a percolative transport pathway at $f_{LNC} = f_p$. Once $f_{LNC}$ increases beyond $f_p$, the electrons can propagate through this percolating network of LNCs, thus opening up a new transport pathway. As $f_{LNC}$ grows past $f_p \rightarrow 1$, the percolating networks densify and the mobility steadily increases toward its value in a monodisperse NC solid, formed only from LNCs. The opening of this new percolating LNC transport channel at $f_{LNC} = f_p$, and its subsequent broadening as $f_{LNC}$ grows past $f_p$ explains the mobility minimum.

**2. Mobility as a function of temperature _T_:** The mobility minimum is considerably smoothed out as the temperature is raised from $T=80K$ to $T=300K$. The mobility at the minimum rises by a bit more than three orders of magnitude. The ratio of the mobilities at the minimum for the two temperatures is consistent with an estimate based on an activated transport across a gap of $\Delta E_{CBM} = 60$ meV. Of course, the precise value of the mobility is further impacted by $f_{LNC}$, and the electron density as well.

We add that more complex behaviors can emerge as a function of temperature. The depth of the mobility minimum is controlled by the $f_{LNC}$ fraction of the LNCs in the BNS and the thermal activation factor exp(-$\Delta/k_BT$), corresponding to an electron hopping from an LNC "trap" to an SNC. In a first approximation, the $\Delta$ energy barrier faced by an electron in an LNC trap can be identified with $\Delta E_{CBM}$. However, for traps that are already occupied, the trap energy $\Delta$ is renormalized to the smaller value of $\Delta_r=\Delta E_{CBM}-E_c$, as long as $\Delta E_{CBM}>E_c$. Next, it is recalled that Kang et al. reported that some experiments can be explained by assuming that the charging energy $E_c$ depends on the temperature to a substantial degree: $E_c=E_c(T)$.[52] In their work, Kang et al. reported a 40-70% increase of $E_c(T)$ as the temperature was raised from $T=20K$ to $80K$. A charging energy $E_c(T)$ that increases with temperature creates a renormalized trap energy $\Delta_r(T)$ that decreases with increasing temperature. This mechanism can explain a smoothing of the mobility minimum with the temperature that is faster than a smoothing driven by temperature-independent energy parameters alone.

Remarkably, the opposite scenario can arise as well. If $\Delta E_{CBM} >> E_c$, then several electrons can occupy each trap. This creates effective traps whose energy spectrum is a ladder with $E_c$ level spacing, and thus $E_c$ plays the role of the renormalized trap energy $\Delta_r=E_c(T)$. In this parameter range, a Kang-type temperature-dependent charging energy creates a renormalized gap $\Delta_r(T)$ that increases with increasing temperature, making the high temperature mobility minimum deeper than the one corresponding to temperature-independent parameters.

The above considerations motivate us to investigate the behavior of the BNS mobility in different parameter ranges, and to use our interaction-renormalized trap model to analyze and interpret our results. This is what we do in the rest of this paper.

**3. Mobility as a function of electron density _n_ and charging energy _E_c_:** The electrostatic effects of adding electrons to a NC are taken into account following the work of Delerue.[47, 48] When an electron is added to a neutral NC at the bottom of the conduction band, the $E_{CBM}$ energy is increased by the one-electron self-energy, $\Sigma$, arising from the charging of the NC, also including the effect of the NC's polarizable host. When a second electron is added, the screened repulsion between the two electrons costs an additional charging energy $E_c$ (denoted as U by Delerue). When the KMC step of HINTS evaluates the probability of a jump from one NC to another NC, the difference between the total initial energy and total final energy is calculated, including these electrostatic energies. The details of the HINTS code are described in the Supporting Information.

When a LNC is unoccupied within a SNC matrix, it is a trap with a depth for an electron. $\Delta$ is given by the difference of the CBM energies of the LNC and the SNC, $\Delta E_{CBM}$, plus the difference of the self-energies $\Sigma_{LNC}-\Sigma_{SNC}$: $\Delta=\Delta E_{CBM}+(\Sigma_{LNC}-\Sigma_{SNC})$. When the trap is already occupied by an electron, the trap energy $\Delta$ gets renormalized by the charging energy $E_c$ to $\Delta_r=\Delta-E_c$. For $E_c<\Delta$, this renormalization transforms the deep traps into shallow traps, making them much less efficient in hindering transport. For $E_c>\Delta$, this renormalization transforms the LNCs from negative-energy traps into positive-energy obstacles.

We explore the effects of varying the charging energy by simulations using $E_{c,LNC}=E_{c,SNC}=E_c=35$, and $E_{c,LNC}=E_{c,SNC}=E_c=125$ meV, in order to explore both relevant regimes of $E_c < \Delta$, and $E_c > \Delta$. The effects of the charging energy $E_c$ are closely interdependent with that of the electron density $n$. We therefore explored the average electron density $n$ sweeping across $n$=0-0.5 e/NC, because most density-dependent phenomena are cyclic with a $n$ period of 1. Therefore, in the complementary range of $n$=0.5-1.0 e/NC, the mobility's behavior is the approximate mirror image across $n$=0.5, and for $n$>1, the entire cycle repeats.

**Figs. 4-5** show simulation results for small and large $E_c$. In both cases, the mobility minimum gets shifted to larger $f_{LNC}$ concentrations as the electron density $n$ grows. The primary reason for this is trap renormalization by electron occupancy. As the electrons are introduced into the sample, they fill up the LNC traps, renormalizing them into shallower traps, or possibly into obstacles. For $n < f_{LNC}$, this reduces the number of deep, unrenormalized traps, which are the primarily suppressants of the mobility. As the electron density exceeds the LNC density, $n$>$f_{LNC}$, a trapped electron population equal to the number of LNCs dynamically renormalizes essentially *all* deep trap LNCs into shallow traps. This leaves the non-trapped excess ($n$-$f_{LNC}$) electrons to move across the BNS that has the same trap density, but which are now renormalized into shallow traps. – In reverse, the ($n$-$f_{LNC}$) density of non-trapped electrons decreases as $f_{LNC}$ increases, thereby decreasing the mobility. Once $f_{LNC}$ exceeds $n$, the mobility keeps decreasing with $f_{LNC}$ until the unrenormalized deep traps percolate. The percolation of the unfilled/unrenormalized deep traps is only reached at $f_{LNC}$ concentrations that exceed $f_p$ by a quantity set by $n$. This explains the mobility minimum moving to higher $f_{LNC}$ concentrations with increasing $n$.

**Fig. 4** shows that when $E_c$ is small (35 meV), and thus $E_c < \Delta$, increasing $n$ shifts the mobility minimum to higher $f_{LNC}$, as well as makes the minima shallower. **Fig. 5** shows that when $E_c$ is large, $E_c > \Delta$, increasing $n$ again shifts the mobility minimum to higher $f_{LNC}$, but without changing its depth. In the case of **Fig. 4**, LNCs with one trapped electron remain energetically capable of trapping additional electrons, as renormalized shallow traps $\Delta_r$. In this situation, increasing $n$ fills an increasing fraction of the LNCs with electrons, thereby decreasing the average trap energy in the NC film and thus resulting in a shallower and shallower mobility minimum. In addition, shallow traps next to deep traps make it easier for the trapped electrons to escape from the deep traps via a two-step process, thus further reducing the depth of the mobility minimum.

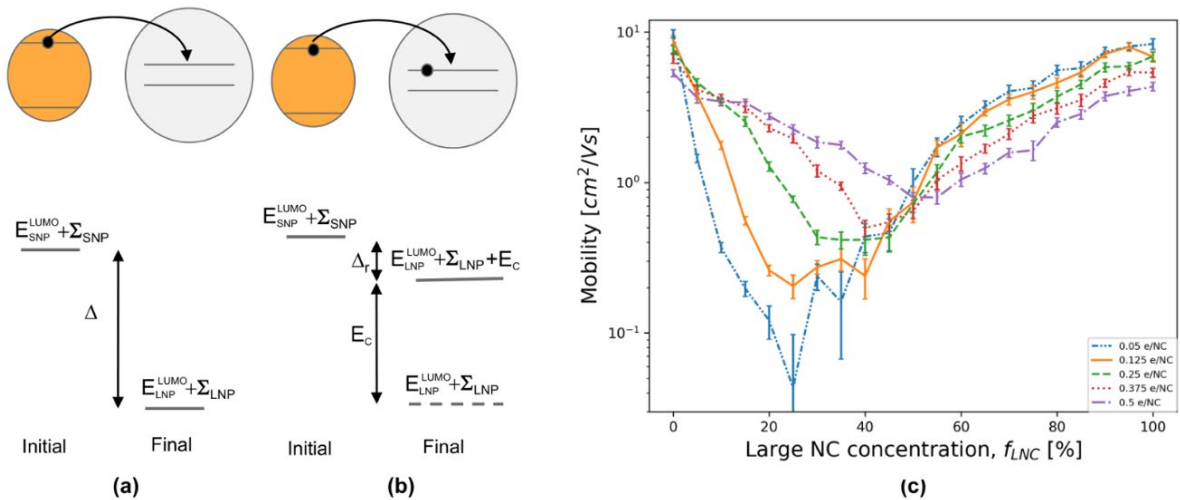

**Fig. 4. Impact of electron concentration when $E_c < \Delta$. Fig. 4a** Energy levels of the initial state and the final state of an electron hopping from an SNC on an unoccupied trap of an LNC. **Fig. 4b** Energy levels of the initial state and the final state of an electron hopping from an SNC on a singly occupied trap of an LNC. The unchanging energy of the electron already on the LNC is not shown expressly for clarity. The dashed level shows the energy without the charging energy $E_c$ for reference. **(c)** Mobility vs. $f_{LNC}$ for electron densities from 0.05 e/NC to 0.5 e/NC at $T$ = 80 K, with $\sigma_{SNC}$ = 0.01 nm and $\sigma_{LNC}$ = 0.08 nm for $E_c<\Delta$. – The mobility at zero LNC concentration is higher than in **Fig. 3**, because the disorder of the SNC diameters, $\sigma_{SNC}$, is four times smaller than in **Fig. 3**.

**Fig. 5** shows that for large $E_c$ (125 meV), since $E_c>\Delta$, the LNCs that have trapped one electron become energetically incapable of trapping an additional electron, and are thus transformed into kinetic obstacles against transport. Therefore, each singly-occupied LNC becomes a lost trap, so increasing $n$ again shifts the mobility minimum to higher $f_{LNC}$, but the average trap energy remains unchanged, and thus the depth of the minimum remains unchanged too, as shown in **Fig. 5**. In addition, increasing $n$ increases the mobility at low $f_{LNC}$ values.

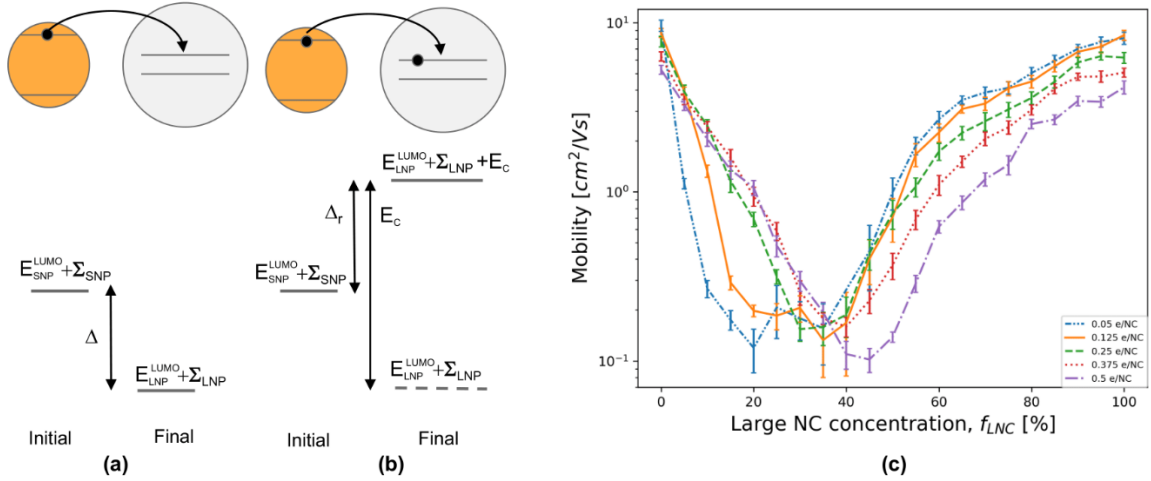

**(a)**      **(b)**      **(c)**

**Fig. 5. Impact of electron concentration when $E_c > \Delta$. Fig. 5a** Energy levels of the initial state and the final state of an electron hopping from an SNC on an unoccupied trap of an LNC. **Fig. 5b** Energy levels of the initial state and the final state of an electron hopping from an SNC on a singly occupied trap of an LNC. The starting energy of the electron already on the LNC is not shown expressly for clarity. The dashed level shows the energy without the charging energy $E_c$ for reference. **(c)** Mobility vs. $f_{LNC}$ for electron densities from 0.05 e/NC to 0.5 e/NC at $T = 80$ K, with $\sigma_{SNC} = 0.01$ nm and $\sigma_{LNC} = 0.08$ nm for $E_c>\Delta$.

### 4. Mobility as a function of size disorder:

We briefly explored the effect of LNC size polydispersity on transport. **Fig. 6** compares mobilities for LNC polydispersities of $\sigma_{LNC} = 0.08$ nm and 0.16 nm. We find that such an increase in polydispersity decreases the depth of the mobility minimum without changing its position. The site energy disorder, induced by the LNC polydispersity, has competing effects. Increasing disorder tends to decrease the mobility on the LNC network itself. However, this same increased variation of the site energies makes some of the LNCs into shallower traps, enhancing the mobility. In the present case, the latter of the two effects seems to be more impactful.

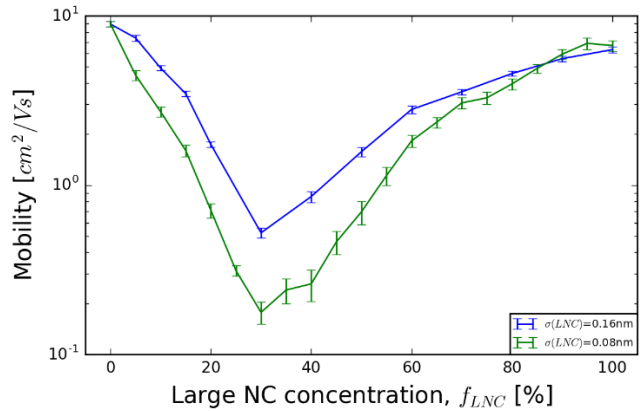

**Fig. 6.** Mobility vs. $f_{LNC}$ for an LNC size polydispersity of $\sigma_{LNC} = 0.08$ nm and $\sigma_{LNC} = 0.16$ nm. Electron density $= 0.25$ $e$/NC; $T = 80$K; SNC polydispersity $\sigma_{SNC} = 0.01$ nm; $E_c = 35$ meV.

**5. Mobility as a function of ligand length:** We simulated transport for ligand lengths of 0.6 nm and 0.8 nm, resulting in NC-NC separations of 1.2 nm and 1.6 nm respectively. The larger inter-NC spacing reduces the entire mobility curve by a multiplicative factor, the ratio of the tunneling factors that depend on the ligand length exponentially. Varying the ligand length did not change the depth or position of the mobility minimum.

**6. Transport Heat Maps:** We visualized the transport through the percolating network by building heat maps that show the time-integrated electron occupancy for each NC during the simulation. At periodic instants we recorded the location of each electron, and overlaid all these images. Thus, an NC appearing darker indicates more electrons spending longer times on that NC. **Figs. 7-8** show heat maps for several different $f_{LNC}$ values, with an electron density equal to 0.25 $e$/NC in a BNS with $E_c < \Delta$. At $f_{LNC} = 0$, the sample consists only of SNCs: the heat map is very homogeneous, and the mobility is high. At $f_{LNC} = 0.05$, the LNCs act as isolated traps, but most electrons can avoid the sparse traps and propagate via the SNC matrix. For $f_{LNC} = 0.10$, the LNC traps start to capture a substantial fraction of the propagating electrons. For $f_{LNC} = 0.15$, most the electrons spend most of their time captured in the LNC traps. These traps are isolated, or form small clusters. At $f_{LNC} = 0.20$, the LNC clusters nearly interconnect. The mobility decreases in the entire range of LNC concentrations from $f_{LNC} = 0$ to $f_{LNC} = 0.20$.

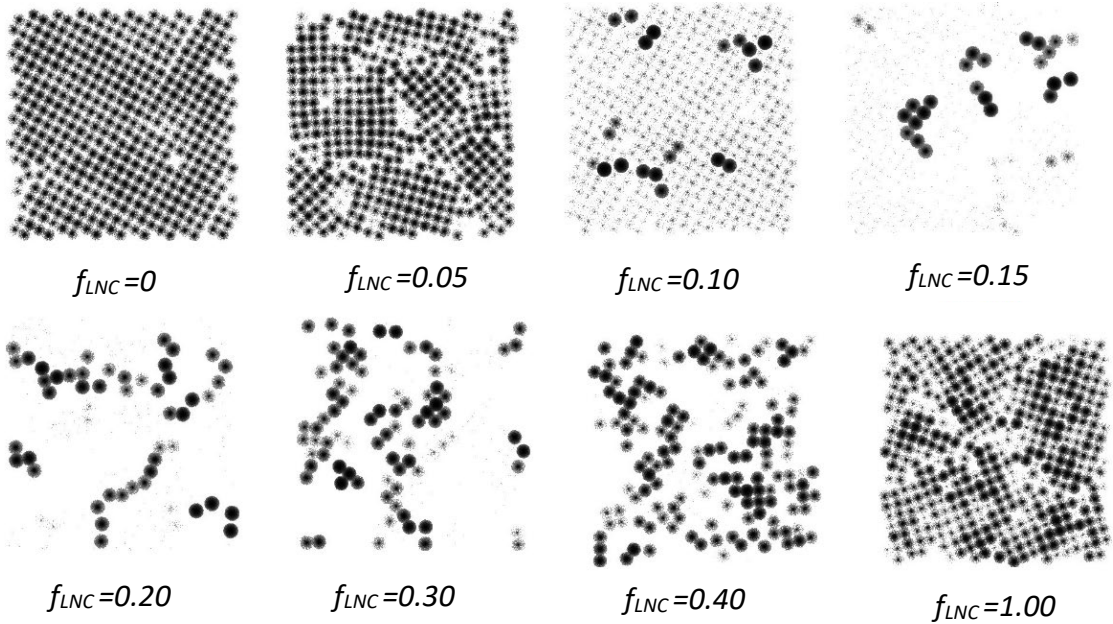

**Figure 7.** Electron occupancy heat maps for a typical BNS for different values of $f_{LNC}$. Shading indicates the time-integrated probability of each NC being occupied by electrons.

At $f_{LNC} = 0.30$, the first sample-spanning LNC clusters appeared. The formation of these percolation networks causes the mobility to begin increasing. At $f_{LNC} = 0.40$, the percolation networks densify, greatly helping the mobility to recover. Finally, at $f_{LNC} = 1.00$, the occupancy heat map becomes quite homogeneous, comparable to the map at $f_{LNC} = 0$. These images give a clear visual support to the physical picture developed above: as $f_{LNC}$ starts to increase from zero, the LNCs serve as traps and thus suppress the mobility. When the LNCs percolate, then the traps suddenly form new transport channels and start increasing the mobility. Therefore, the mobility forms a minimum at the percolation threshold $f_p$, modified by the electron density $n$. As $f_{LNC}$ grows towards $f_{LNC} = 1.00$, the transport can again flow through the entire NC matrix.

**Fig. 8** shows the effect of temperature on the heat maps of a typical BNS. Electron transport pathways are dominated by the deep traps at $T = 80$ K, giving rise to a very uneven heat map. At $T=300$ K, thermal energy is reasonably effective at freeing the trapped electrons, so the electron transport is more homogeneous throughout the sample. This translates to a higher mobility at higher temperatures, which is consistent with the smoothing out of the mobility minimum at $T=300$K, reported above in **Fig. 3**.

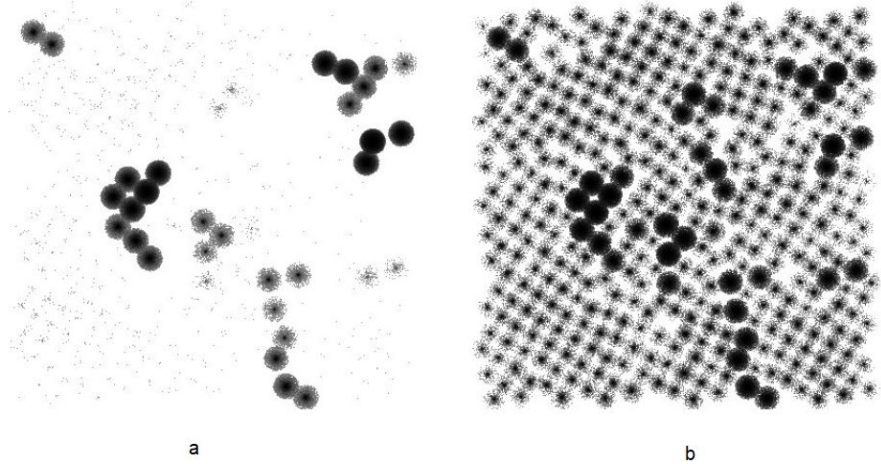

a                                   b

**Figure 8.** Temperature dependence of the electron occupancy maps for a typical BNS at (a) $T = 80$ K and (b) $T = 300$ K.

## Discussion and Conclusions

We simulated and fabricated field-effect transistors (FETs) made from a mixture of PbSe NCs with diameters of 6.5 nm and 5.1 nm, thereby forming a binary nanocrystal solid, BNS. We used our Hierarchical Nanoparticle Transport Simulator to model the transport in these BNSs, and study the impact of several factors on this transport. The BNS mobility exhibited a minimum at a large-NC-fraction $f_{LNC}=0.25$. The mobility minimum was deep at T=80K and partially smoothed at T=300K. We developed the following physical picture to account for this behavior. As the LNC fraction $f_{LNC}$ within the SNC matrix starts growing from zero, the few LNCs act as deep traps for the electrons traversing the SNC matrix. Increasing the $f_{LNC}$ concentration of these traps decreases the mobility. As the increasing $f_{LNC}$ reaches the percolation threshold $f_{LNC}=f_p$, the LNCs form sample-spanning networks that enable electrons to traverse the entire BNS via low-energy, low disorder LNC pathways. The opening of the new transport channel through these percolating LNC pathways leads to the recovery of the mobility as $f_{LNC}$ grows past $f_p$. Therefore, the electron mobility exhibits a pronounced minimum as a function of $f_{LNC}$ at $f_{LNC}=f_p$. We have studied the effect of temperature, electron density, charging energy, ligand length, and disorder on the mobility minimum. To account for all trends, we have proposed that capturing an electron renormalizes a deep trap LNCs into either a shallow trap or a kinetic obstacle, depending on the value of the charging energy $E_c$ relative to the NC energy difference $\Delta$. A central prediction of this model is that the position of the mobility minimum shifts to a larger LNC fraction $f_{LNC}>f_p$ as the electron density increases, but its depth is modified differently depending on whether $E_c<\Delta$, or $E_c>\Delta$. Finally, we verified our expectations and physical picture by constructing and analyzing heat maps of the mobile electrons in the simulated BNS.

## Methods

We have developed HINTS, a hierarchical, multi-level Kinetic Monte Carlo theory of transport in binary nanocrystalline solids. The main levels of our method were described very briefly in the main text in order to provide context for the presentation of the results and their discussion. The details of our methods are provided in the Supporting Information, available online. The experimental methods of fabricating the modeled binary nanocrystalline solids were also mentioned in the main text, and are explained in detail in the Supporting Information online.

## Associated Content

Supporting Information accompanies this article and is available free of charge.

## Author Information

### Corresponding Author

*Davis Unruh - orcid.org/0000-0001-6992-9617; dgunruh@ucdavis.edu

### Author Contributions

An early version of the code was developed by G.Z. in a collaborative effort. The code was substantially developed and extended by L.Q. The simulations were performed by L.Q. and D.U. All authors contributed to the development of the main modules of the method. The study was conceived and directed by G.Z. All authors contributed to the discussion of the results. The manuscript was primarily written by G.Z., with all authors contributing.

### Funding Sources

The work was supported by the NSF National Science Foundation under award DMR-2005210.

### Notes

The authors declare no competing financial interest.

## Acknowledgements

The authors thank Nicholas Brawand, Ian Carbone, Giulia Galli, Victor Klimov, Art Nozik, Boris Shklovskii, Marton Voros, and especially Matt Law for many useful discussions.

# Supporting Information for: Percolative Charge Transport in Binary Nanocrystal Solids

*Luman Qu, Davis Unruh\*, and Gergely T. Zimanyi*

Physics Department, University of California, Davis, Davis, California 95616, United States

*dgunruh@ucdavis.edu

## Experimental Methods

*Chemicals*. Lead oxide (PbO, 99.999%), selenium (99.99%), oleic acid (OA, tech. grade, 90%), diphenylphosphine (DPP, 98%), trioctylphosphine (TOP, tech. grade, >90%), 1-octadecene (ODE, 90%), 1,2-ethanedithiol (EDT, >98%), trimethylaluminum (97%), and anhydrous solvents were purchased from Aldrich and used as received. Millipore water was degassed with three freeze-pump-thaw cycles before loading into the atomic layer deposition (ALD) system.

*Nanocrystal Synthesis*. PbSe nanocrystals (NCs) were synthesized and purified using standard air-free techniques. 1.5 g of PbO, 5 g of oleic acid, and 10 g of 1-octadecene was stirred in a three-neck flask at 180°C for 1 hr. 9.5 milliliters of a 1 $M$ solution of trioctylphosphine selenide containing 0.2 milliliters of diphenylphosphine was then rapidly injected into the hot solution. To control the NC size, the NCs were allowed to grow at 160°C for preselected times (several minutes). The reaction was then quenched with a water bath and 15 milliliters of anhydrous hexane. The NCs were purified by three rounds of dispersion/precipitation in hexane/ethanol and stored as a powder in a glovebox for later use.

*NC Film Deposition*. NC solutions of various LNC number fraction were prepared by suspending appropriate amounts of LNC and SNC powder in dry hexane at a total concentration of 2 mg mL$^{-1}$. A mechanical dip coater (DC Multi-4, Nima Technology) installed inside of a glovebox was used to prepare NC films via layer-by-layer deposition [**Error! Reference source not found.**]. Briefly, substrates (glass or prepatterned FET substrates, cleaned by sonication in isopropanol and dried under $N_2$ flow) were alternately dipped into the NC solution and a 1 mM solution of 1,2-ethanedithiol (EDT) in dry acetonitrile. We fabricated films with thicknesses of 30 ± 5 nm for field-effect transistors and 50-100 nm for optical measurements. The fraction and spatial distribution of large NCs in each film type was measured by SEM imaging (FEI Magellan 400) of dip-coated, oleate-capped NC monolayers prior to ligand exchange with EDT.

*Atomic Layer Deposition Infilling*. The NC transistors were infilled and overcoated with ~20 nm of amorphous $Al_2O_3$ deposited in a homemade cold-wall traveling wave ALD system within a glovebox from trimethylaluminum and water at a substrate temperature of 54°C and an operating base pressure of about 88 mTorr. Pulse and purge times were 9 ms and 60 s, respectively.

*Other Characterization*. Transmission electron microscopy (TEM) imaging was performed on a Philips CM20 operating at 200 kV. Optical absorption spectra were acquired with a PerkinElmer Lambda 950 spectrophotometer operating in transmission mode.

## Theoretical Methods

*Electronic NC Structure*: As described in the main text, the energy of the conduction band minimum, $E_{CBM}$, is calculated using the $\boldsymbol{k \cdot p}$ method of Kang & Wise.[Error! Reference source not found.] We apply a rigid shift to align the infinite diameter limit of the conduction band edge with the bulk work function of PbSe.

The on-site self-charging energy $E_c$ is the energy cost that needs to be paid to load each electron onto a given nanocrystal. In general, the accounting is defined by:

$$E_C = n\left(\Sigma + \frac{n-1}{2}E_C\right)$$

where $n$ is the number of electrons that will be on the NP after the electron is added. $\Sigma$ is the energy that needs to be paid upon the load of the first charge onto the neutral NC, while $E_C$ is the extra energy it takes to load each additional charge (due to interactions with other charges as well as their induced image charge).

As mentioned in the main text, in our model both $\Sigma$ and $E_C$ are calculated using the single NP empirical-perturbative hybrid calculations of Delerue,[1] consistent with single NP *ab initio* calculations[1] and the self-capacitive term used by experimental reports.[Error! Reference source not found.] This method gives the results:

$$\Sigma = \frac{q^2}{8\pi\epsilon_0 R}\left(\frac{1}{\epsilon_{\text{solid}}} - \frac{1}{\epsilon_{NC}}\right) + 0.47\frac{q^2}{4\pi\epsilon_0 R}\left(\frac{\epsilon_{NC} - \epsilon_{\text{solid}}}{\epsilon_{NC} + \epsilon_{\text{solid}}}\right)$$

and

$$E_C = \frac{q^2}{4\pi\epsilon_0 R}\left(\frac{1}{\epsilon_{\text{solid}}} + 0.79\frac{1}{\epsilon_{NC}}\right)$$

where $R$ is the radius of the NC, $\epsilon_{\text{solid}}$ is the dielectric constant of the medium, and $\epsilon_{NC}$ is the dielectric constant inside the NC.

For the dielectric constant inside the NC, we assume that it will be the bulk high frequency dielectric constant of PbSe, taken to be 22.0. To properly treat the dielectric constant of the medium, we account for both the organic ligand shell of the NP, as well as the presence of neighboring NPs. We assume that the ligands themselves have a dielectric constant of 2.0. The dielectric constant of the entire BNS is then calculated using the Maxwell-Garnett (MG) effective medium approximation:

$$\epsilon_{\text{solid}} = \epsilon_{\text{ligand}}\frac{\epsilon_{NC}(1 + \kappa f) - \epsilon_{\text{ligand}}(\kappa f - \kappa)}{\epsilon_{\text{ligand}}(\kappa + f) + \epsilon_{NC}(1 - f)}$$

where $\kappa$ is 2 for spherical NPs, and $f$ is the filling factor.

*Electron Transition Rates*: We have adopted Miller-Abrahams single-phonon assisted hopping as our framework for calculating these transition probabilities. An alternate approach also included as an option in our code is the Marcus theory of multi-phonon activated transitions, but Miller-Abrahams typically yields the more accurate results. Under Miller-Abrahams, the transition probabilities are given by:

$$\Gamma_{i \rightarrow j} = \nu\beta_{ij}\exp\left(-\frac{\Delta E_{ij}}{k_b T}\right)$$

where the attempt frequency $\nu$ is chosen to be $10^{12}\text{s}^{-1}$ to match experimental data. $\Delta E_{ij}$ is the total energy difference associated with an electron transitioning from $NC_i$ to $NC_j$: $\Delta E_{ij} = \Delta E_{ij}^{\text{band}} + \Delta E_{ij}^{\text{voltage}} + \Delta E_{ij}^{\text{charging}}$, where $\Delta E_{ij}^{\text{band}}$ is the contribution from the single-particle band energies of the NCs, $\Delta E_{ij}^{\text{charging}}$ is the contribution from the NC charging energy, and $\Delta E_{ij}^{\text{voltage}}$ is the energy difference due to the applied electric field. Specifically, $\Delta E_{ij}^{\text{voltage}} = qV(z_j - z_i)/L_z$ where $V$ is the voltage difference between the two sides of the BNS in the $\hat{z}$ direction (the transport direction), $L_z$ is the entire length of the BNS in the $\hat{z}$ direction, and $(z_j - z_i)$ is the z distance between the two NPs.

Here it should also be noted that $\beta_{ij}$ is the tunneling amplitude, calculated using the WKB approximation:

$$\beta_{ij} = \exp\left(-2\Delta x \sqrt{\frac{2m^*\left(E_{\text{vac}} - E_{\text{tunneling}}\right)}{\hbar^2}}\right)$$

Here $\Delta x$ is the minimum surface separation of the NCs, set to be twice the ligand length. $m^*$ is the effective mass of electrons in the tunneling medium, approximated as $.05m_e$, the effective mass of electrons in bulk PbSe. It is noted that $m^*$ was estimated to be $.3m_e$ in NCs. $E_{\text{vac}}$ is the vacuum energy level, set to be zero as all other energy levels are defined relative to the vacuum. The energy of the well that the electron resides in, $E_{\text{tunneling}}$, is taken to be equal to $\frac{E_i + E_j}{2}$, the average of the initial and final energy states of the hopping transition. This is the approach of Chandler & Nelson.**Error! Reference source not found.**

*Kinetic Monte Carlo Framework*: Monte Carlo (MC) algorithms are a class of numerical algorithms which incorporate random numbers in attempts to accurately simulate real-world problems. Broadly, this is implemented by calculating the probability of specific events, and then executing them if a random number between 0 and 1 falls in that specified range.

A Kinetic Monte Carlo algorithm is a specific MC which seeks to accurately simulate the evolution in time of physical systems. It does so by propagating the physical system forwards step by step, executing a single event and evolving time accordingly at every step. In this manner, KMC simulations mimic explicitly the evolution of real-world systems.

Our Kinetic Monte Carlo simulator is implemented according to the BKL algorithm.[3] At each step, the KMC tabulates the probabilities of all possible events occurring, and then chooses which one to execute by choosing a random number $r_1$ between 0 and 1. Time is then evolved by choosing another random number $r_2$ (also between 0 and 1). The specifics are as follows.

Once the simulation is initialized (and all charges have been placed at random into the network), the time-evolution starts by determining the transition rates $\Gamma$ of all possible events, which will be locally updated after each algorithm step. In each step, the event $j$ is executed using the random number $r_1$ for which the following equation is satisfied:

$$\sum_{i=1}^{j-1} \Gamma_i < r_1\Gamma_{\text{sum}} < \sum_{i=j+1}^{N} \Gamma_i$$

where

$$\Gamma_{\text{sum}} = \sum_{i=1}^{N} \Gamma_i$$

After this execution, time is evolved according to the second random number, $r_2$:

$$\Delta t = -\frac{\ln(r_2)}{\sum_{ij} \Gamma_{j \to i}}$$

Finally, the list of all possible events is updated to reflect the current state of the system, and the rate table is updated accordingly.

Our simulations typically run for 500,000 events, well into the steady state. Data is only collected after the transients have dissipated.

Mobility is measured in our simulation as:

$$\mu = \frac{(\text{electron collected at drain electrode}) * l}{(\text{total number of electrons}) * t * E}$$

where $l$ the length of the simulation box normal to the electrodes, $t$ is the simulation time, and $E$ is the applied electric field. The harvested charge is the total charge of the electrons which have reached the drain electrode. Our simulations use periodic boundary conditions, so these electrons will pass through the drain electrode to the source electrode to be reintroduced to the sample. Any electron that travels in the opposite direction gives a negative addition to the current.

---